\documentclass[aps,prb,reprint,amsmath,amssymb,superscriptaddress,showpacs,floatfix]{revtex4-1}

\usepackage{graphicx}
\usepackage{dcolumn}
\usepackage{bm}
\usepackage[colorlinks, allcolors=blue]{hyperref}
\usepackage[usenames, dvipsnames]{color}

\begin{document}

\title{Quantum phases of frustrated 2-leg spin-1/2 ladders with skewed rungs}

\author{Geetanjali Giri}
\thanks{First and second authors have contributed equally ~~~~}
\email{geetgiri1@gmail.com}
\affiliation{Solid State and Structural Chemistry Unit, Indian Institute of Science, Bangalore 560012, India}

\author{Dayasindhu Dey}
\email{dayasindhu.dey@bose.res.in}
\affiliation{S. N. Bose National Centre for Basic Sciences, Block - JD, Sector - III, Salt Lake, Kolkata - 700098, India}

\author{Manoranjan Kumar}
\email{manoranjan.kumar@bose.res.in}
\affiliation{S. N. Bose National Centre for Basic Sciences, Block - JD, Sector - III, Salt Lake, Kolkata - 700098, India}

\author{S. Ramasesha}
\email{ramasesh@sscu.iisc.ernet.in}
\affiliation{Solid State and Structural Chemistry Unit, Indian Institute of Science, Bangalore 560012, India}

\author{Zolt\'an G. Soos}
\email{soos@princeton.edu}
\affiliation{Department of Chemistry, Princeton University, Princeton, New Jersey 08544, USA}

\date{\today}

\begin{abstract}
The quantum phases of 2-leg spin-1/2 ladders with skewed rungs are obtained 
using exact diagonalization of systems with up to 26 spins and by density matrix renormalization 
group calculations to 500 spins. The ladders have isotropic antiferromagnetic 
(AF) exchange $J_2 > 0$ between first neighbors  in   the   legs,  variable 
isotropic AF exchange $J_1$ between some first neighbors in different legs, and 
an unpaired spin per odd-membered ring when $J_1 \gg J_2$. Ladders with skewed 
rungs and variable $J_1$ have frustrated AF interactions leading to multiple 
quantum phases: AF at small $J_1$, either F or AF at large $J_1$, as well as 
bond-order-wave phases or reentrant AF (singlet) phases at intermediate $J_1$. 
\end{abstract}


\maketitle

\section{\label{sec:intro}Introduction}
Many different kinds of spin chains have been studied over years. 
Bethe~\cite{bethe31} and Hulthen~\cite{hulthen38} obtained the ground state (GS) of 
the linear spin-1/2 Heisenberg antiferromagnet (HAF). The HAF is a prototypical 
gapless system with isotropic exchange $J > 0$ between neighbors and is closely 
realized in crystals that contain $S = 1/2$ transition metal ions or organic 
molecular ions. Haldane~\cite{haldane83} pointed out that the spin-1 HAF is a 
gapped system as was soon confirmed both experimentally and numerically. The 
$J_1$-$J_2$ model, Eq.~\ref{eq:hsb} below, has isotropic $J_1$, $J_2$ between 
first and second neighbors, respectively. It has been extensively studied in 
connection with frustrated interactions.~\cite{chitra95,*sahoo2014,ckm69b,lecheminant2004,hase2004,
sudan2009,*dmitriev2008,furukawa2012} Depending on the ratio $J_1/J_2$, the 
system has gapless phases with quasi-long-range order and gapped dimer or 
incommensurate phases.~\cite{soos-jpcm-2016} Even when limited to spin-1/2 chains, 
there is considerable freedom in the number and range of exchange interactions, 
in the choice of isotropic (Heisenberg), anisotropic or antisymmetric exchange, 
or spin ladders with two or more parallel chains. Sandvik~\cite{sandvik2010} 
has reviewed numerical approaches to spin chains and ladders. 

There is considerable interest in the spin gap of the fermionic chains for applications
in singlet fission.~\cite{Michl10,suryo} The spin gap can be reduced by raising the  energy of the
GS with respect to the triplet state, by  introducing kinetic frustration,
via fused odd membered rings. Thomas et al.~\cite{thomas2012} studied the fused
azulenes both in spin models with isotropic antiferromagnetic exchange interaction
and in the Pariser-Parr-Pople (PPP) model for $\pi$-electrons. 
The azulene molecule (C$_{10}$H$_8$) has fused 5 and 7  membered rings that, when
fused into a polymer, define the unit cell of a 2-leg ladder with skewed rungs.
To their surprise Thomas et al.~\cite{thomas2012} found that the spin 
of the GS of the system increased with system size in both models. 

Recent interest in spin chains has focused on exotic 
quantum phases~\cite{sudan2009, *dmitriev2008,furukawa2012}
and the possibility of multiferroic materials.~\cite{cheong2007,katsura2005,ren2012,kobayashi2012,thomas2012}
However, so far there has not been a systematic study to improve the rate at which the GS spin 
increases or to  unravel the reason behind cascading spin of the GS with increasing system size. 
This study is aimed at understanding the magnetic GS of fused azulene in particular and fused 
frustrated ring systems in general.

\begin{figure}[b]
\includegraphics[width=0.9\columnwidth]{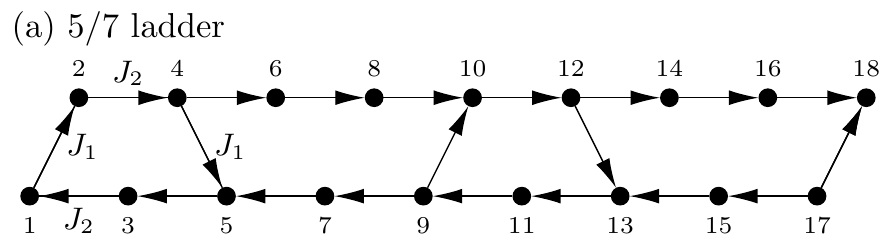}
\includegraphics[width=0.9\columnwidth]{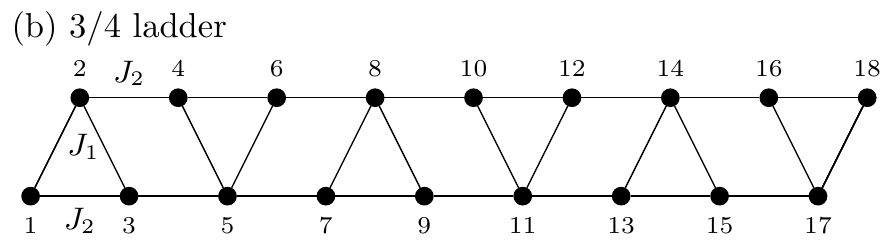}
\caption{\label{fig1}Schematic representation of the 5/7 ladder (top) and 3/4 
ladder (bottom) with isotropic exchange $J_1$ in rungs and $J_2$ between neighbors in 
legs. The arrows are discussed in Sec.~\ref{sec5}.}
\end{figure}

We obtain in this paper the quantum phases of frustrated 2-leg spin-1/2 ladders 
with skewed rungs. Such ladders have not been studied previously to the best of 
our knowledge. The ladders can be viewed as generalizations of the $J_1$-$J_2$ 
model: Isotropic $J_2$ between second neighbors corresponds to HAFs on legs of 
odd and even numbered sites; isotropic $J_1$ between first neighbors corresponds 
to two zig-zag rungs per site. The ladders we discuss have fewer $J_1$ rungs as 
illustrated in Fig.~\ref{fig1}. 
The rungs of the 5/7 ladder correspond to 
fused azulenes while the rungs of the 3/4 ladder define alternating fused 
3 and 4-membered rings. We consider ladders with variable number of spins, variable $J_1 > 0$ and 
constant $J_2 = 1$. Ladders with skewed rungs may have equivalent legs 
(Fig.~\ref{fig1}b) or inequivalent legs (Fig.~\ref{fig1}a) in addition to at least one  
frustrated odd-membered ring per unit cell.

The legs of all ladders are conveniently numbered in Fig.~\ref{fig1} as odd or 
even integers. Each leg is a HAF with isotropic $J_2$ between neighbors and either 
periodic boundary conditions (PBC) or open boundary conditions (OBC). Skewed rungs 
with isotropic $J_1$ connect adjacent spins in this numbering. The $J_1$-$J_2$ model 
is the 3/3 ladder
\begin{equation}
H_{3/3} = J_1 \sum_{r} \vec{S}_r \cdot \vec{S}_{r+1} 
        + J_2 \sum_{r} \vec{S}_r \cdot \vec{S}_{r+2}
\label{eq:hsb}
\end{equation}
The 3/3 ladder has one spin per unit cell and the largest possible number of 
skewed rungs. The GS is a singlet, $S_G = 0$, in the entire 
sector $J_1, J_2 > 0$ that includes spin liquid phase and gapped dimer or 
incommensurate phases.~\cite{soos-jpcm-2016} All ladders in this paper have identical 
legs with $J_2 = 1$ in Eq.~\ref{eq:hsb} and different numbers of $J_1$ rungs. The 
conventional 2-leg ladder is $H_{4/4}$ with one $J_1$ rung, two spins per unit cell, 
and inversion symmetry at the middle of every rung. The GS is a singlet~\cite{barnes93,azzouz94} 
with a finite singlet-triplet gap for $J_1 > 0$. Ladders with skewed rungs 
instead have inversion symmetry at some sites in one or both legs.

Quite generally, large $J_1 > J_2$ localizes a spin on every odd-membered ring. 
The sign of the effective exchange between spins determines whether the GS is 
ultimately F or AF in the thermodynamic limit. The total spin $S$ is conserved 
and ranges from $S_G(J_1,N) = 0$ in AF ladders to $S_G(J_1,N) \propto N$ in F 
ladders with exchange $J_1 > J_2 = 1$. Increasing $J_1$ generates a variety of 
quantum phases. In addition to F or AF (singlet) phases, some ladders support bond order 
waves (BOWs) or reentrant AF phases as a function of $J_1/J_2 > 0$.

The present study was motivated by the 5/7 ladder in Fig.~\ref{fig1}a. The GS 
is a singlet at $J_1 = 1$ for $N < 50$ and $S_G(1,N)$ increases slowly with 
system size.~\cite{thomas2012} The origin of F interactions in systems with 
purely AF exchange and the thermodynamic limit are difficult to assess for the 
large unit cell of eight spins. We address both issues in ladders with smaller 
unit cells. It turns out that the evolution of $S_G(J_1,N)$ with $J_1$ or system 
size is quite variable in ladders with skewed rungs and leads to multiple quantum 
phases. The ladders share a common qualitative feature, however: The GS is always 
a singlet at small $J_1$ and changes of $S_G(J_1,N)$ typically cluster around 
$J_1 \sim 2$. In the approximation of singlet pairing of adjacent sites, or 
Kekul\'e valence bond diagrams, three $J_2$ bonds at small $J_1$ are converted 
into two $J_1$ bonds and two unpaired spins at large $J_1$. Accurate treatment 
of ladders generates other phases at intermediate $J_1$ before reaching an AF 
or F phase at large $J_1$.

The paper is organized as follows. Numerical methods are summarized in Sec.~\ref{sec2}. 
The quantum phases of the 3/4 and 5/5 ladders are presented in Sec.~\ref{sec3}. 
The 3/4 ladder has a first-order AF to F transition around $J_{1c} = 1.58$ while the 
5/5 ladder has a narrow BOW phase and a singlet GS over the entire range. The 3/5 
ladder in Sec.~\ref{sec4} has multiple quantum phases at intermediate $J_1 \sim 2$ 
and frustrated effective exchanges leading to a singlet GS at large $J_1$. The 5/7 
ladder in Sec.~\ref{sec5} also has multiple quantum phases at intermediate $J_1$, 
including a reentrant AF phase, and a F phase for $J_1 > 2.4$ with an unpaired spin 
per ring. The discussion in Sec.~\ref{sec6} summarizes the pattern of F or AF 
phases of other 2-leg ladders with skewed rungs.
\section{\label{sec2}Numerical methods}
We use exact diagonalization (ED) for ladders up to 26 spins. The relevant 24-spin 
ladders with PBC have an integral number of unit cells and sectors with inversion 
symmetry $\sigma$ at some sites. The extra rung gives 26 spins in OBC ladders. 
Matrix elements, correlation functions and excited states are also computed.

\begin{figure}[b]
\includegraphics[width=0.95\columnwidth]{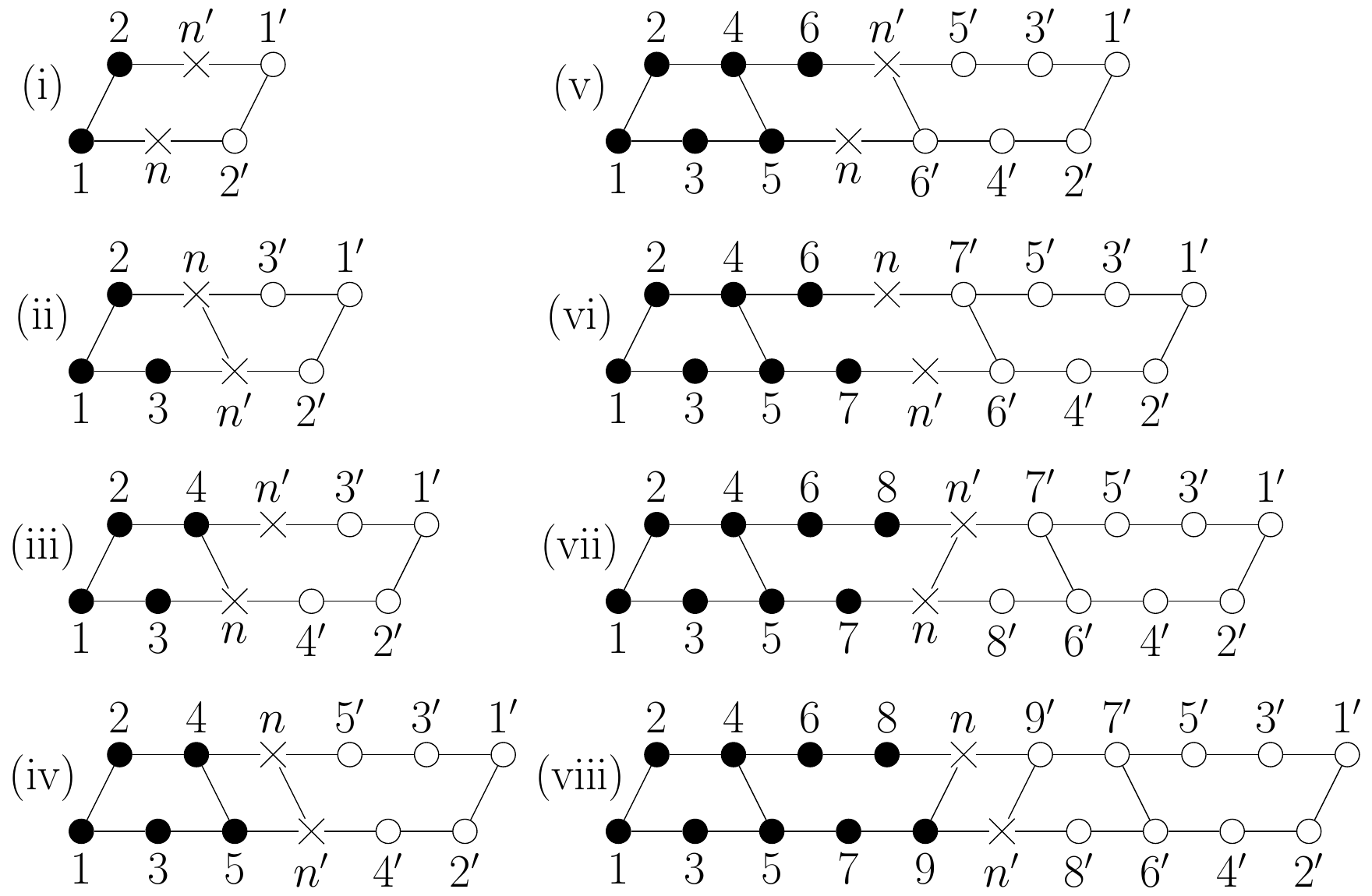}
\caption{\label{fig2} Schematics of the DMRG steps to grow a 5/7 ladder.
The filled (open) circles are the left (right) of the ladder and the crosses are
the new sites introduced at each steps. While the left (right) is treated as 
the system the right (left) works as the environment. Other skewed ladders in this study are also similarly constructed.} 
\end{figure}

We use the Density Matrix Renormalization Group (DMRG) 
technique~\cite{white-prl92,*white-prb93} to obtain the low energy states of larger OBC systems. The DMRG scheme for building skewed ladders is similar to building regular ladders and proceeds by adding two new sites at a time, starting with a ring of four sites. The schematic for building a 5/7 ladder is shown in Fig.~\ref{fig2}.  Since OBC ladders are asymmetric about the middle in general, we have used asymmetric DMRG in which we keep track of the left and right systems separately and have carried out 10 sweeps of the finite system DMRG. 
In the  worst case, the truncation error in our calculation is below $10^{-11}$ by 
keeping up to $m=500$ eigenvectors  corresponding to the highest eigenvalues of the density 
matrix. The  error in  total energies are estimated to be less than $10^{-5}\%$
which leads to  uncertainty in energy gaps of less than $10^{-4}$. 
Comparable DMRG accuracy is discussed elsewhere for other kinds of spin 
chains.~\cite{schollwock2005,karen2006,chitra95,mk2010,ddpbc2016} The largest ladders 
studied in this paper have almost 500 sites.   

The total spin $S$ is always conserved and is explicitly taken into account 
in the valence bond (VB) basis,~\cite{ramasesha84,soos89} as is inversion 
symmetry at sites in PBC ladders. The spin $S_G(J_1,N)$ of the absolute GS 
is obtained by comparing the lowest energy in sectors with different $S$
\begin{equation}
\Gamma_S(N) = E_0(S, N) - E_0(0, N).
\label{eq:gap}
\end{equation}
We have $S_G = 0$ when $\Gamma_S > 0$, level crossing when $\Gamma_S = 0$, 
and $S_G > 0$ given by the largest $S_G$ for which $\Gamma_S < 0$. Similarly, 
the gap $\Gamma_{\sigma}$ at fixed $S_G$ is to the lowest state with reversed 
inversion symmetry 
\begin{equation}
\Gamma_{\sigma} = E_0(S_G, \sigma = -1, N) - E_0(S_G, \sigma = 1, N).
\label{eq:gma-sig}
\end{equation}
The GS is even when $\Gamma_{\sigma} > 0$, odd when $\Gamma_{\sigma} < 0$, and 
doubly degenerate when $\Gamma_{\sigma} = 0$. The singlet-triplet gap $\Gamma_T(N)$ 
is the excitation energy to the lowest triplet state in systems whose GS 
is a singlet, $S_G = 0$.

DMRG gives accurate results for the low-energy states of long ladders. The $z$ 
component of the total spin, $S^z$, is conserved and exploiting this conservation
is straightforward. The highest $S^z$ value for which $E_0(S^z) - E_0(0)$ is zero
defines the spin $S_G$ of the GS. $S_G(J_1,N)$ is inferred from the energies 
of the lowest $S^z$ states. When $S_G > 0$, the $(2S_G + 1)$ 
Zeeman components are degenerate. Hence $E_0(0,N)$ is an excited state in the 
$S^z = 0$ sector, as are the $S^z = 0$ components of states with $S < S_G$. It 
follows that DMRG gives $E_0(S^z,N) - E_0(S^z - 1,N) = 0$ when $S_G$ increases 
from $S_G - 1$ to $S_G$. In practice, we start with $S^z = 0$ and increase it 
by integer steps; $S_G$ is reached when the $E_0$ at $S^z = S_G + 1$ is higher 
than at $S^z = S_G$.~\cite{thomas2012} Although DMRG does not specify inversion symmetry, the GS 
is degenerate within the numerical accuracy when $\Gamma_{\sigma}(N) = 0$.

\section{\label{sec3}The 3/4 and 5/5 Ladders}
\begin{figure}
\includegraphics[width=0.9\columnwidth]{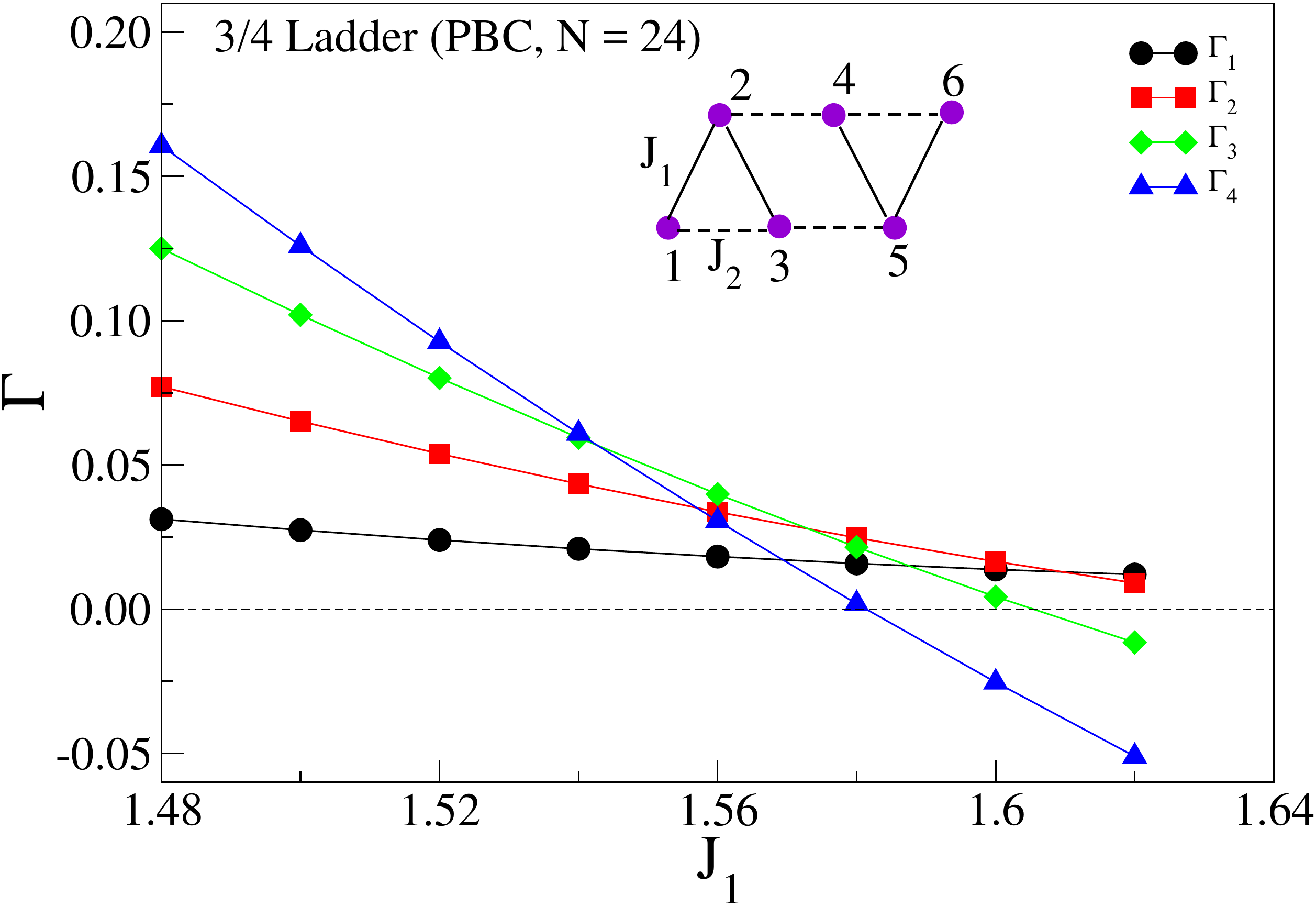}
\caption{\label{fig3}Energy differences $\Gamma_S$ in Eq.~\ref{eq:gap} in sectors 
with total spin $S$ as a function of $J_1$ for a 3/4 ladder with PBC and 24 spins; 
inset: two unit cells of the 3/4 ladder under PBC.}
\end{figure}

The 3/4 ladder in Fig.~\ref{fig1}b has 2/3 as many $J_1$ rungs as the $J_1$-$J_2$ 
ladder. It has two consecutive $J_1$ followed by a missing rung. The Hamiltonian is
\begin{equation}
H_{3/4} = J_1 \sum_{r} \vec{S}_{3r-1} \cdot \left( \vec{S}_{3r-2} + \vec{S}_{3r} \right)
        + J_2 \sum_{r} \vec{S}_r \cdot \vec{S}_{r+2}.
\label{eq:h34}
\end{equation}
The PBC ladder has three spins per unit cell and inversion symmetry at the apex 
of isosceles triangles with sides $J_1$ and base $J_2$. As shown in Fig.~\ref{fig3}, 
the PBC ladder of $N = 8n = 24$ spins has a singlet GS for $J_1 < J_{1c}(N)$ and 
$S_G = 4n/3$ for $J_1 > J_{1c}$. Each triangle has an unpaired spin and the jump 
from $S_G = 0$ to 4 at $J_{1c} = 1.58$ indicates a ferromagnetic effective exchange between 
triangles. The inset shows $J_2$ exchanges at sites 2,4 and 3,5 of adjacent triangles. 

In addition to the spin $S_G$ as a function of $J_1 \ge 0$, $J_2 = 1$, we compute 
the spin density $\rho_j$ at site $j$ and spin correlation functions $C(j,k)$ as 
the expectation values
\begin{equation}
\begin{aligned}
\rho_j &= 2 \langle G | S_j^z | G \rangle \\
C(j, k) &= \langle G | \vec{S}_j \cdot \vec{S}_k | G \rangle.
\end{aligned}
\label{eq:corl}
\end{equation}
Spin densities vanish identically in singlet states. The PBC ladder has two 
first-neighbor spin correlations in Table~\ref{tab1} with $k = j \pm 1$ that 
vanish at $J_1 = 0$ and two second-neighbor correlations with $k = j \pm 2$, 
one of which changes sign with increasing $J_1$. 
\begin{table}
\caption{\label{tab1}First and second neighbor spin correlation functions 
$C(j,k)$ of the 3/4 ladder with PBC and N = 24 spins. Refer to Fig.~\ref{fig1}
for site numbers.}
\begin{ruledtabular}
\begin{tabular}{ c  c  c  c  c }
$J_1$ & $C(1,2), \, C(2,3)$ & $C(3,4)$ & $C(1,3)$ & $C(2,4), \, C(3,5)$ \\ \hline
0    &  0      & 0      & -0.4489 & -0.4489 \\
1    & -0.1201 & 0.0972 & -0.3587 & -0.4255 \\
1.5  & -0.3004 & 0.1801 & -0.0363 & -0.3522 \\
1.56 & -0.3215 & 0.1975 &  0.0036 & -0.3400 \\
2    & -0.4480 & 0.2272 &  0.2350 & -0.2497 \\
5    & -0.4923 & 0.1655 &  0.2486 & -0.1308 \\
20   & -0.4995 & 0.1253 &  0.2499 & -0.0737 \\
40   & -0.4999 & 0.1182 &  0.2500 & -0.0646 \\ 
\end{tabular}
\end{ruledtabular}
\end{table}

Spins in different legs are uncorrelated at $J_1 = 0$. The second neighbor 
correlations at $J_1 = 0$ are $(1/4 - \ln 2) =  -0.44315$ in the thermodynamic 
limit. Finite size effects are fairly small at $N = 24$. The sign of $C(1,3)$ 
changes near the jump from $S_G = 0$ to 4. The triangles at large $J_1$ have 
a doublet GS with $C(1,2) = C(2,3) = -0.5$ and $C(1,3) = 0.25$ that are almost 
reached at $J_1 = 5$. The limiting spin densities are $\rho_a = -1/3$ at the 
apex and $\rho_b = 2/3$ at each base, which gives an unpaired spin per triangle. 
The spin densities at $J_1 = 5$ and 20 are, respectively, $\rho_a = -0.4269$ 
and $-0.3578$, and $\rho_b = 0.7135$ and $0.6789$. They converge more slowly 
with $J_1$ than spin correlations. At large $J_1$ we have $C(2,4) = C(3,5) = 
\rho_a \rho_b/4 = -1/18 = -0.0556$. The limit now requires $J_1 > 40$ due to 
the slow evolution of spin densities.

\begin{figure}
\includegraphics[width=0.9\columnwidth]{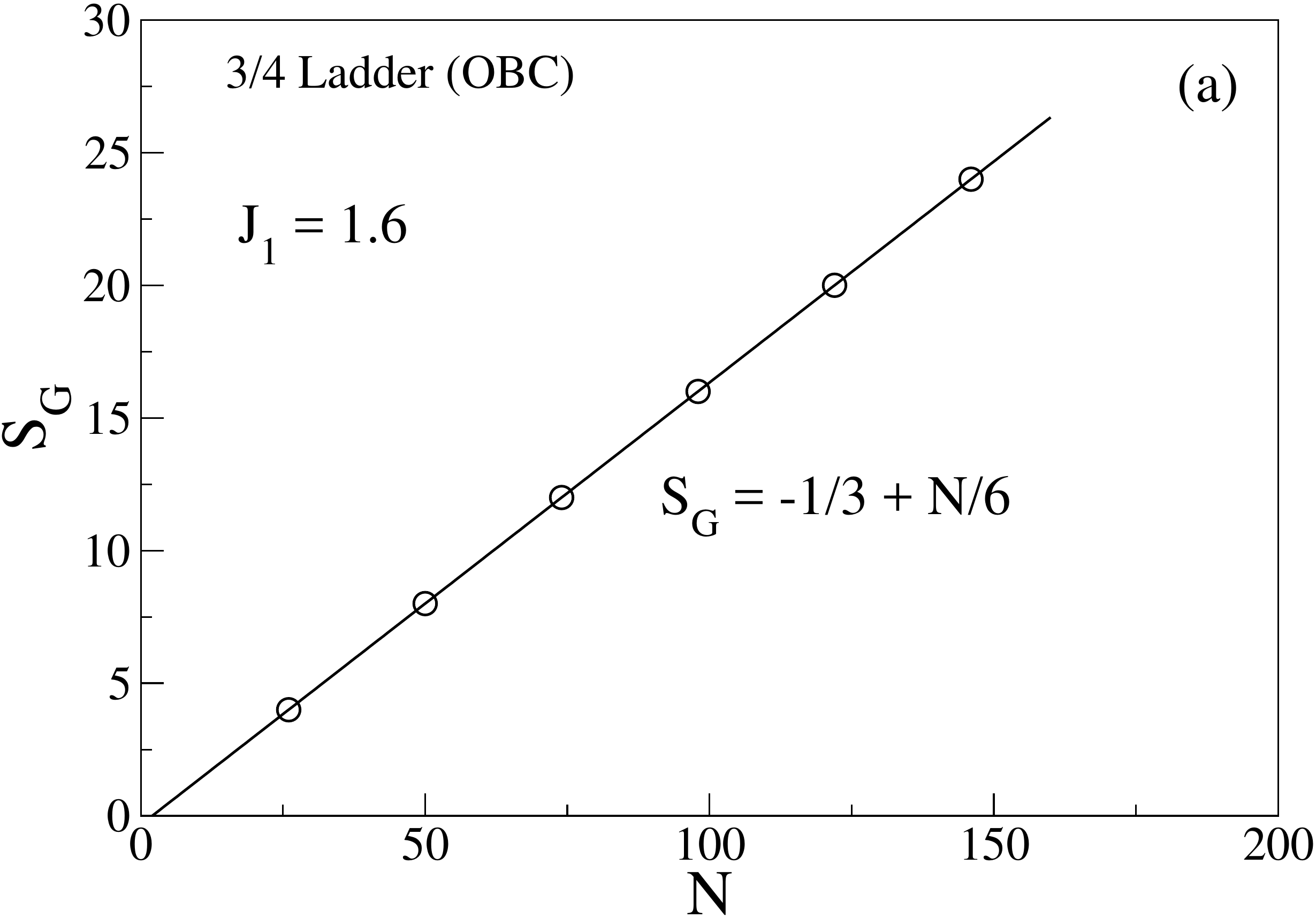}
\includegraphics[width=0.9\columnwidth]{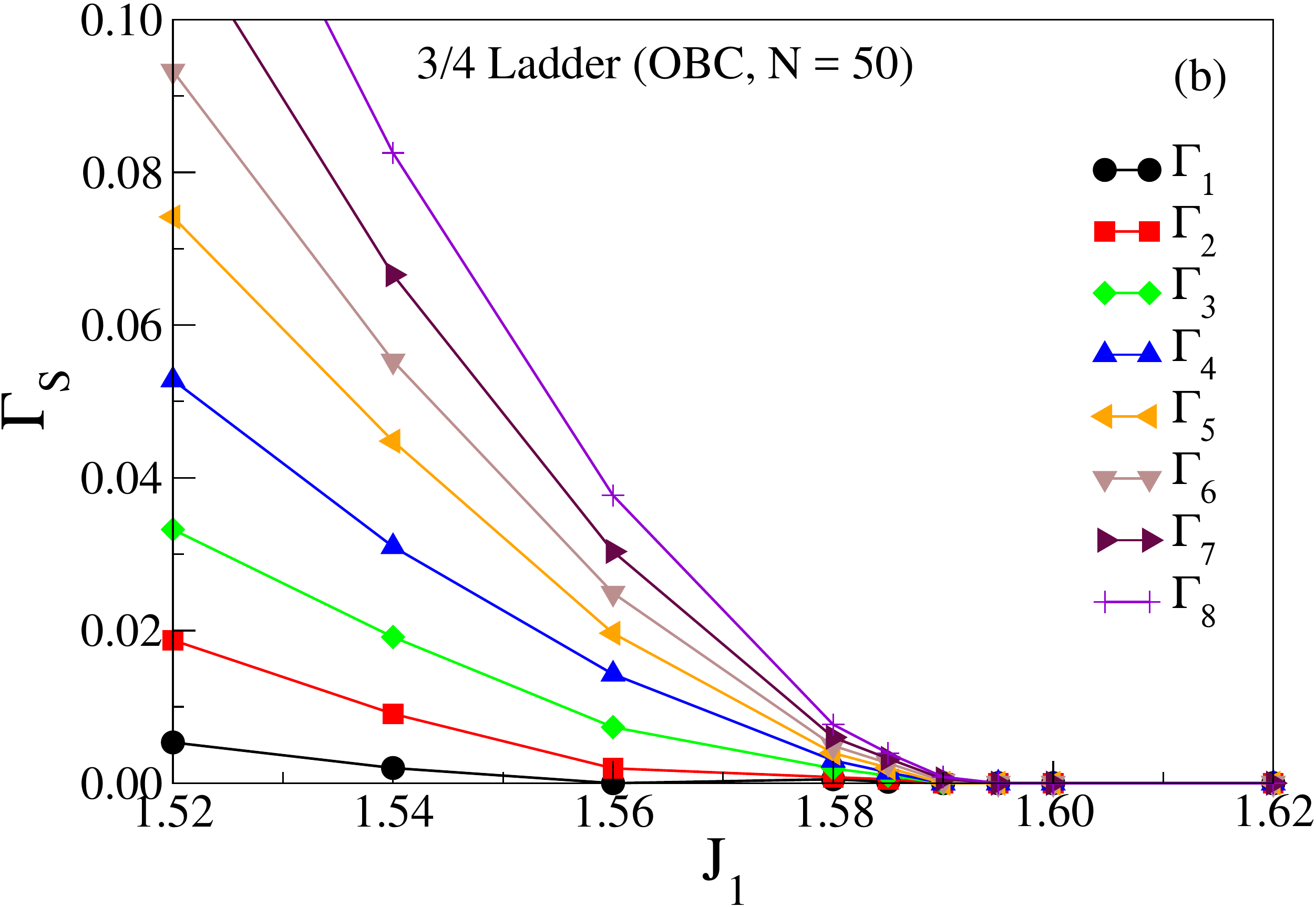}
\caption{\label{fig4}(a) The GS spin $S_G$ of 3/4 ladders with OBC and 
$J_1 = 1.6$ as a function of system size. The F phase has one spin per triangle. 
(b) The energy difference $\Gamma_S$ between the GS in sectors with $S^z = S_G$ to 0 
as a function of $J_1$ for a 3/4 ladder with OBC and 50 spins.}
\end{figure}

The OBC ladders have $N = 8n + 2$ spins and a $J_1$ rung between sites 
$8n + 1$ and $8n + 2$. The GS still has $S_G = 4n/3$ and one unpaired spin per 
triangle for $J_1 \ge 1.6$. DMRG results in Fig.~\ref{fig4}, upper panel, have 
increasing $S_G$ and indicate a F phase in the thermodynamic limit with one 
spin per three-membered ring. $S_G$ changes rapidly but sequentially with 
increasing $J_1$ in OBC ladders, from $S_G = 0$ to 4 for 24 spins. DMRG for 
50 spins in Fig.~\ref{fig4}, lower panel, shows that $S_G$ jumps from 0 to 1 
at $J_1 = 1.56$ and reaches the expected $S_G = 8$ by $J_1 = 1.59$. 

The ferromagnetism of 3/4 ladders for $J_1 > 1.58$ is an example of the McConnell 
mechanism~\cite{mcconnell63} with AF exchange (here $J_2$) between sites with 
positive and negative spin densities to obtain a F interaction. The effective F 
exchange between adjacent triangles is $J_{\mathrm{eff}} = 2 J_2 \rho_a \rho_b$ 
and goes to $-4/9$ for $J_2 = 1$ in the limit $J_1 \gg 1$. The McConnell idea has 
been generalized to inorganic as well as organic radicals~\cite{kollmar93} with 
delocalized electrons and has been realized experimentally at low temperature in 
oligomers of spin-1/2 radicals.~\cite{izuoka87}

\begin{figure}
\includegraphics[width=0.9\columnwidth]{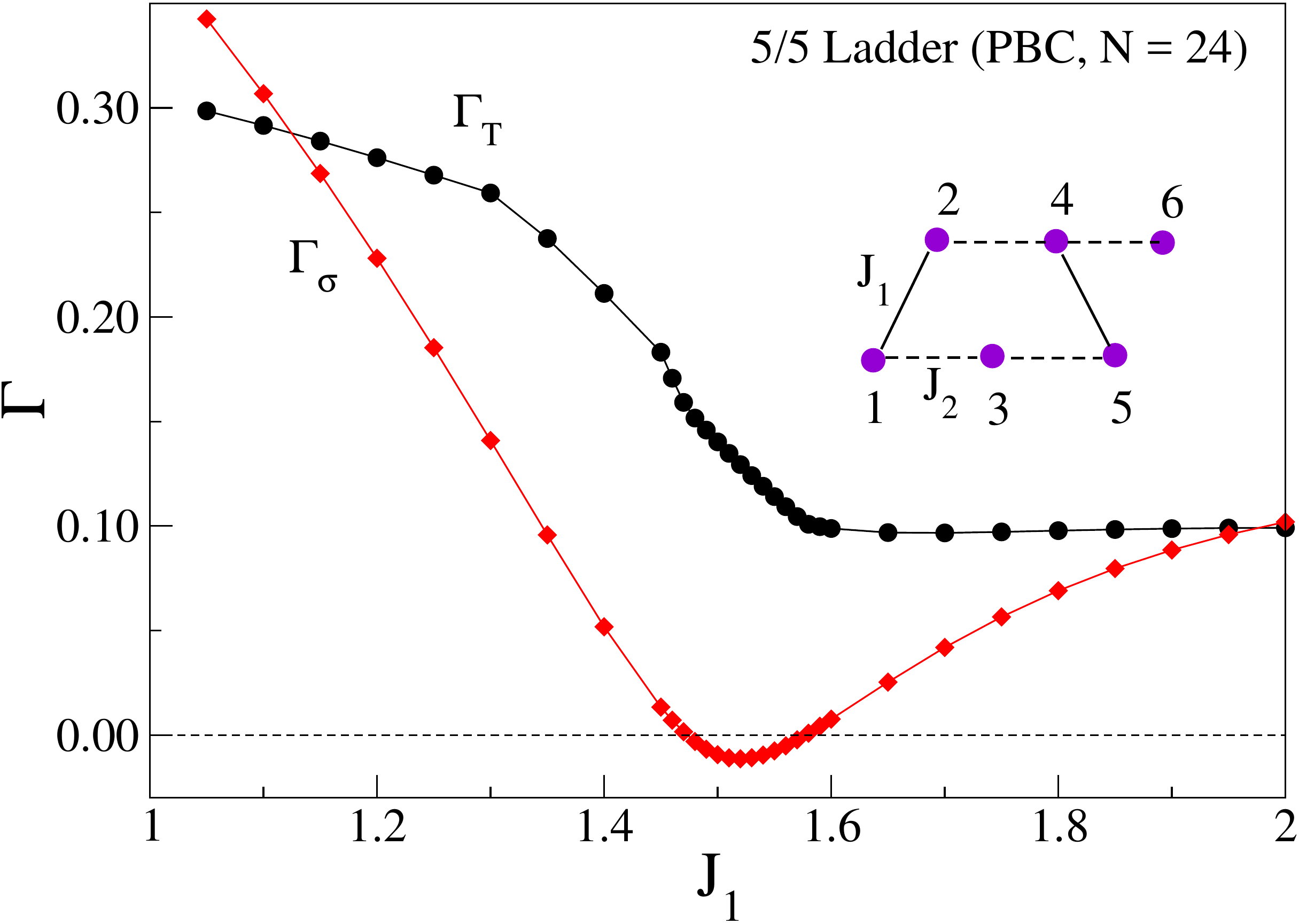}
\caption{\label{fig5}Excitation energies of a 5/5 ladder with PBC and 24 spins: 
$\Gamma_T$ to the lowest triplet state and $\Gamma_{\sigma}$ to the lowest 
singlet with opposite inversion symmetry to the singlet GS; inset: two unit cells.}
\end{figure}

The 5/5 ladder (Fig.~\ref{fig5}, inset) has two exchanges $J_1$ per six $J_2$.
The Hamiltonian is
\begin{equation}
H_{5/5} = J_1 \sum_{r} \vec{S}_{3r-2} \cdot  \vec{S}_{3r-1}
        + J_2 \sum_{r} \vec{S}_r \cdot \vec{S}_{r+2}.
\label{eq:h55}
\end{equation}
There are half as many rungs as in the 3/4 ladder. The GS is a singlet, $S_G = 0$, 
over the entire range $J_1 > 0$. Large $J_1$ localizes a spin at sites $3r$, the 
only sites without a $J_1$ rung. Second order perturbation theory gives an AF 
effective exchange $J_{\mathrm{eff}} \sim J_2^2 /J_1$ between adjacent spin-1/2 
at sites $3r$. The system is paramagnetic in the limit of infinite $J_1$. The 
thermodynamic limit at large $J_1$ is a HAF for localized spins at sites $3r$; 
the gapless singlet phase has quasi-long-range order.

ED results for $N = 24$ with PBC are entirely consistent with these expectations. 
Table~\ref{tab:55corl} lists spin correlation functions at sites $3r$. Spins in 
different legs have $C(j,k) = 0$ at $J_1 = 0$ and weak correlations at $J_1 = 1$. 
Near neighbor $C(j,k)$ of the HAF are known analytically~\cite{shiro05} in the thermodynamic limit. $C(3,9)$ 
is between third neighbors in a leg where the $J_1 = 0$ result is $C_3 =  -0.15074$. 
Increasing $J_1$ reverses the signs of $C(j,k)$ in the same leg and makes more 
negative the $C(j,k)$ in different legs. For $J_1 = 5$ or 40, the $C(j,k)$ in 
Table~\ref{tab:55corl} can be compared to the  first through fourth neighbors of an 
8-site HAF ring: $C(1,2) = -0.4564$, $C(1,3) = 0.1958$, $C(1,4) = -0.1890$ and $C(1,5) = 0.1491$. 
The HAF results in the thermodynamic limit~\cite{shiro05} are almost the same at $C_1 = -0.44315$
and $C_2 = 0.18204$. The 8-site correlations are larger as expected than
$C_3$ and $C_4$ (= 0.10396). The effective Hamiltonian at large $J_1$ is an HAF with spins at sites $3r$.
\begin{table}
\caption{\label{tab:55corl}Spin correlation functions $C(j,k)$ at sites $3r$ of 
the 5/5 ladder with PBC and $N = 24$. Refer to Fig.~\ref{fig5}, inset, for site numbers.}
\begin{ruledtabular}
\begin{tabular}{ c  c  c  c  c }
$J_1$ & $C(3,6)$ &  $C(3,9)$ & $C(3,12)$ & $C(3,15)$ \\ \hline
1	 &  -0.0176  & -0.1294    &   -0.0098   &  0.0917   \\
1.5	 &  -0.1489  & -0.004     &   -0.0051   &  -0.0018  \\
2	 &  -0.2977  &  0.111     &   -0.1028   &  0.0809   \\
5	 &  -0.426   &  0.1796    &   -0.1706   &  0.1339   \\
40	 &  -0.4559  &  0.1948    &   -0.1874   &  0.1478   \\
\end{tabular}
\end{ruledtabular}
\end{table}

The energy gaps $\Gamma_T$ in Eq.~\ref{eq:gap} to the lowest triplet and 
$\Gamma_{\sigma}$ in Eq.~\ref{eq:gma-sig} to the
lowest singlet with reversed inversion symmetry are shown in Fig.~\ref{fig5}
as functions of $J_1$. The GS is doubly degenerate when $\Gamma_{\sigma}(J_1) = 0$ 
at a point or over an interval. The ladder has $\Gamma_{\sigma} = 0$ at 
$J_1 = 1.47$ and $1.58$ in Fig.~\ref{fig5}. The singlet GS is odd under 
inversion between these points. The excited states cross at $J_1 = 1.13$ and 
$2.0$ where $\Gamma_T = \Gamma_{\sigma}$.

The GS degeneracy indicates broken inversion symmetry and a bond-order-wave phase. 
The BOW phase, either dimer or incommensurate, of the $J_1$-$J_2$ 
model has recently been studied in detail,~\cite{soos-jpcm-2016} and we discuss 
below the BOW phase of 3/5 ladders. The 5/5 ladder of 24 spins contains 8 unit 
cells. The 8 spin $J_1$-$J_2$  model has a doubly degenerate GS with 
$\Gamma_{\sigma}= 0$ at two values $J_1 > 0$, and $n/4$ degeneracies with 
$J_1 > 0$ in larger systems. Additional GS degeneracies are likely in longer 
5/5 ladders but the larger unit cell poses numerical difficulties. Likewise, the 
excited state degeneracies $\Gamma_{\sigma}= \Gamma_T$ at $J_1 = 1.13$ and $2.0$ 
in Fig.~\ref{fig5} give a first estimate of the BOW phase in the thermodynamic limit.

\section{\label{sec4}The 3/5 ladder}
\begin{figure}
\includegraphics[width=0.9\columnwidth]{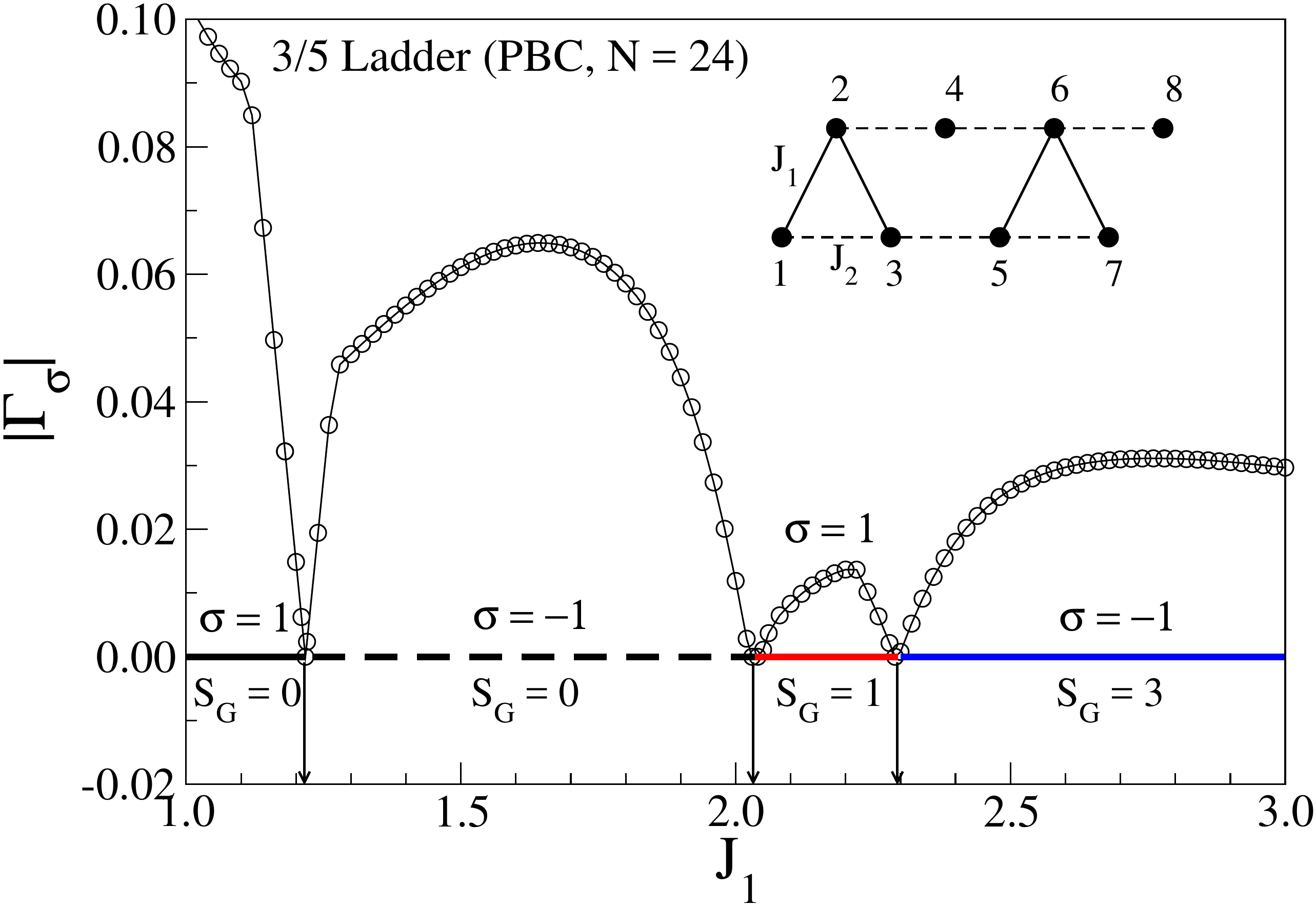}
\caption{\label{fig6}Excitation $\Gamma_{\sigma}$ in Eq.~\ref{eq:gma-sig} of
the 3/5 ladder with PBC and 24 spins; inset: two unit cells. The quantum phases
have spin $S_G$ and inversion symmetry $\sigma$.}
\end{figure}

The 3/5 ladder (Fig.~\ref{fig6}, inset) has four spins per unit cells and is 
the first example of a ladder with inequivalent legs. The Hamiltonian is 
\begin{equation}
H_{3/5} = J_1 \sum_{r} \vec{S}_{4r-2} \cdot \left( \vec{S}_{4r-3} + \vec{S}_{4r-1} \right)
        + J_2 \sum_{r} \vec{S}_r \cdot \vec{S}_{r+2}.
\label{eq:h35}
\end{equation}
There is one rung at each odd-numbered site and two or zero rungs at alternate  
even-numbered sites. In PBC ladders the sites $4r - 2$ and $4r$ are inversion 
centers at the apices of triangles and pentagons, respectively. Inequivalent 
legs are additional frustration beyond odd-membered rings and may be responsible 
for the multiple quantum phases of 3/5 and 5/7 ladders.

The singlet GS of the $N = 24$ PBC system is even under inversion ($\sigma = 1$)
up to $J_1 = 1.22$, as shown in Fig.~\ref{fig6}, where it becomes odd
($\sigma = -1$) and remains odd until $J_1 = 2.03$ where it switches to $S_G = 1$.
The F state of this system has $S_G = 3$, which is reached at $J_1 = 2.30$.
Inversion symmetry in the singlet GS is broken at $J_1 = 1.44$ for $N = 16$.

At $J_1 = 1.22$, the GS with $\sigma = \pm 1$ are degenerate (Fig.~\ref{fig6}). The plus and minus linear combinations, 
$| \pm 1\rangle = \left[| G(\sigma = 1)\rangle \pm | G(\sigma = -1)\rangle \right]/\sqrt 2$, are BOWs with broken inversion symmetry and doubled unit cells. The BOW amplitude of the bond between sites $j$ and $k$ is half of the magnitude of the difference $\langle 1| \vec{S}_j \cdot \vec{S}_k | 1 \rangle - \langle -1 | \vec{S}_j \cdot \vec{S}_k | -1 \rangle$. This gives
\begin{equation}
B(j,k) = \langle G(\sigma = 1) | \vec{S}_j \cdot \vec{S}_k | G(\sigma = -1) \rangle
\label{eq:mat-elm}
\end{equation}
In general, $B(j,k)$ is finite for sites that are not related by inversion.
Since inversion does not interchange the legs, $B(j,k)$ is finite for $j$ 
and $k$ on different legs, i.e. $j + k$ is an odd integer. For spins in the same leg, some sites are related
by inversion and have $B(j,k) = 0$; for example, $B(1,3) = B(1,7) = 0$;
$B(2,6) = B(2,10) = 0$. Table~\ref{tab2} lists $B(j,k)$ up to third neighbors
for $N = 16$ and 24 spins. The largest amplitude is for first neighbors in
the even-numbered leg. The second largest is $B(5,4) = -B(3,4)$ at sites
without a $J_1$ rung; $B(1,2) = -B(2,3)$ at sites connected by$J_1$ is smaller
but decreases more slowly with system size.

\begin{table}
\caption{\label{tab2}BOW amplitudes $B(j,k) =  -B(j',k') > 0$ in Eq.~\ref{eq:mat-elm}
up to third neighbors in 3/5 ladders with PBC and $J_1 = 1.22 \, (N = 24)$ or
$1.44 \, (N = 16)$. Only finite $B(j,k)$ are shown.}
\begin{ruledtabular}
\begin{tabular}{ c  c  c }
$j,k \qquad       j',k'$ & $B(j,k)  \quad N = 16$ & $B(j,k)  \quad N = 24$ \\ \hline
$1,2 \qquad       2,3  $ & $0.03992$ & $0.03776$ \\
$5,4 \qquad       4,3  $ & $0.10378$ & $0.06689$ \\
$2,4 \qquad       4,6  $ & $0.22675$ & $0.18063$ \\
$7,4 \qquad       4,1  $ & $0.05334$ & $0.02206$ \\
$2,5 \qquad       3,6  $ & $0.02068$ & $0.02087$ \\
\end{tabular}
\end{ruledtabular}
\end{table}

\begin{figure}
\includegraphics[width=0.9\columnwidth]{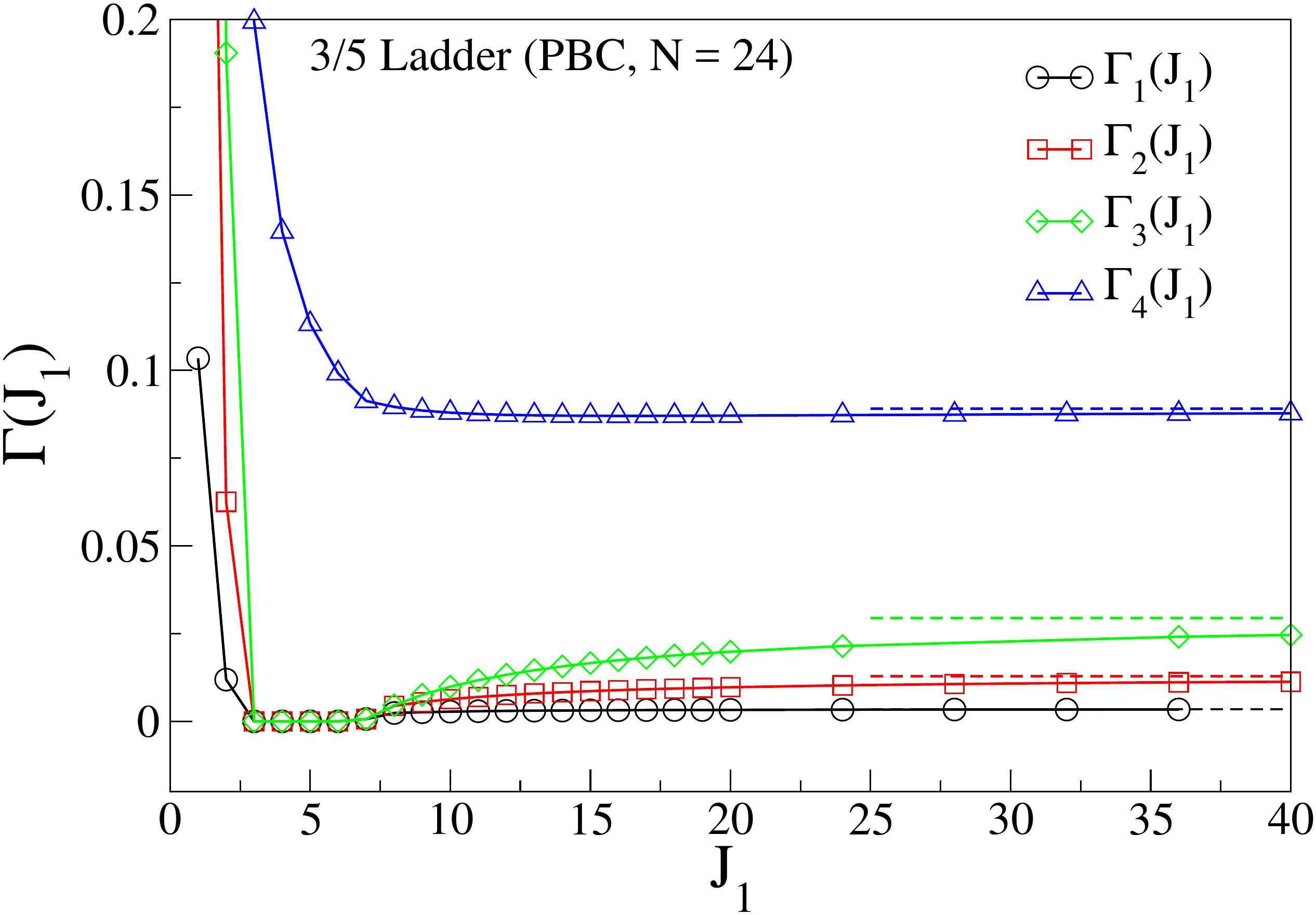}
\caption{\label{fig7} Excitation $\Gamma_S$ in Eq.~\ref{eq:gap} of a 3/5 ladder with 
PBC and 24 spins; the GS is a singlet for $J_1 > 6.87$. The dashed lines are the 
$J_1 \to \infty$ excitations of $H_{\mathrm{eff}}(24)$ in Eq.~\ref{eq:heff-35}.}
\end{figure}

Next we consider 3/5 ladders at large $J_1$. For 24 spins, we find $S_G(J_1,N) = 3$ 
from  $J_1 = 2.30$ in Fig.~\ref{fig6} to $J_1 = 6.87$ in Fig.~\ref{fig7}, where the 
GS reverts to $S_G = 0$. The GS of the 16 spin system is $S_G = 2$ over almost exactly 
the same $J_1$ range. The F phase with an unpaired spin per triangle is limited to 
intermediate $J_1$. We are not aware of another spin-1/2 chain with an intermediate 
F phase between two AF phases.

Selected spin correlation functions $C(j,k)$ of 3/5 ladders are listed in 
Table~\ref{tab3}. PBC ladders have two different first neighbor correlations and 
three different second neighbor correlations. The $J_1$ dependencies follow the 
discussion above of 3/4 ladders, especially with respect to triangles, and the 
$J_1 = 0$ systems are of course identical. $C(1,3)$ changes sign around $J_1 \sim 2$ 
and is close to $0.25$ by $J_1 = 10$, where $C(1,2) = C(2,3)$ is close to $-0.50$. 
The effective AF exchange between spins $3$ and $5$ in adjacent triangles is 
$\rho_b^2 = 4/9$ in the limit $J_1 \gg 1$; $C(3,5)$ approaches a finite 
constant at $J_1 > 40$.
\begin{table}
\caption{\label{tab3}First and second neighbor spin correlation functions 
$C(j,k)$ of the 3/5 ladder with PBC and $N = 24$ spins.}
\begin{ruledtabular}
\begin{tabular}{ c  c  c  c  c  c}
$J_1$  &    $C(1,2),$ & $C(3,4),$ &     $C(1,3) $ &     $C(3,5) $ &     $C(2,4),$ \\ 
       &    $C(2,3) $ & $C(4,5) $ &               &               &     $C(4,6) $ \\ \hline
$0   $ &    $ 0	    $ & $0	$ & 	$-0.4489$ &	$-0.4489$ &	$-0.4489$ \\ 
$1   $ &    $-0.0795$ & $-0.0032$ &	$-0.2709$ &	$-0.5758$ &	$-0.4255$ \\
$1.5 $ &    $-0.2221$ & $0.0041	$ &	$-0.1394$ &	$-0.5218$ &	$-0.3430$ \\
$2   $ &    $-0.2983$ & $0.0185	$ &	$-0.0399$ &	$-0.4484$ &	$-0.2950$ \\
$2.5 $ &    $-0.4534$ & $0.0676	$ &	$ 0.2078$ &	$-0.3314$ &	$-0.1443$ \\
$5   $ &    $-0.4900$ & $0.0583	$ &	$ 0.2431$ &	$-0.2553$ &	$-0.0795$ \\
$10  $ &    $-0.4978$ & $0.0508	$ &	$ 0.2486$ &	$-0.2215$ &	$-0.0472$ \\
$20  $ &    $-0.4995$ & $0.0456	$ &	$ 0.2497$ &	$-0.2062$ &	$-0.0335$ \\
$40  $ &    $-0.4999$ & $0.0428	$ &	$ 0.2499$ &	$-0.1985$ &	$-0.0267$ \\
\end{tabular}
\end{ruledtabular}
\end{table}

The GS of triangles is a doublet at large $J_1$ when other degrees of freedom are 
frozen out. The 3/4 ladder reduces to F exchange between adjacent triangles. The 
3/5 ladder in this limit has AF exchange $j_2 = 4/9$ between the bases of adjacent 
triangles and F exchange $j_1 = -1/3$ between apices (sites $4r - 2$) and spins at 
sites $4r$. A 3/5 ladder of $N = 4n$ spins reduces to a PBC system of $N/2$ spins 
with an effective Hamiltonian that becomes exact as $J_1 \to \infty$.
\begin{equation}
H_{\mathrm{eff}}(N) = j_1 \sum_{r=1}^{N/2} \vec{S}_{2r} \cdot \vec{S}_{2r+2}
  + j_2 \sum_{r=1}^{N/4} \vec{S}_{4r-2} \cdot \vec{S}_{4r+2}
\label{eq:heff-35}
\end{equation}
Eq.~\ref{eq:heff-35} is defined on the even-numbered leg with exchange $j_1$ 
between first neighbors and $j_2$ between sites that correspond to adjacent 
triangles. When $j_2 > 0$, $H_{\mathrm{eff}}$ is a frustrated spin chain for 
either sign of $j_1$. If we set $j_1 = 0$, the first neighbor spin correlations 
in the $j_2$ chain is $C(3,5) = (1 - 4 \ln 2)/9 = -0.19695$, which is close 
to the $N = 24$, $J_1 = 40$ entry in Table~\ref{tab3}. The dashed lines 
Fig.~\ref{fig7} are excitations of $H_{\mathrm{eff}}(24)$ with $j_1 = -1/3$ 
and $j_2 = 4/9$. The six spins at sites $4r$ are weakly coupled and frustrated; 
they account for small gaps up to $\Gamma_3$. The same reasoning explains why 
$H_{\mathrm{eff}}(16)$ has small gaps $\Gamma_1$, $\Gamma_2$ and a larger gap 
$\Gamma_3$. The effective Hamiltonian returns equally quantitative excitations 
for $N = 12$, 16 and 20 spins. 

$H_{\mathrm{eff}}$ has been discussed previously 
by Hamada et al.~\cite{hamada88} in the context of a frustrated spin chain 
related to the $J_1$-$J_2$ model. The GS with $j_2 > 0$ is F for $j_1 \le -2 j_2$ 
and a singlet otherwise.~\cite{hamada88} The thermodynamic limit for $j_1/j_2 = -3/4$ 
is a gapless AF phase with a non-degenerate singlet GS and quasi-long-range 
spin correlations.

\section{\label{sec5}The 5/7 ladder}
The 5/7 ladder (Fig.~\ref{fig1}a) has eight spins per unit cell and two $J_1$ 
rungs per eight exchanges $J_2 = 1$. The Hamiltonian is
\begin{eqnarray}
H_{5/7} &=& J_1 \sum_r \left( \vec{S}_{8r-4} \cdot \vec{S}_{8r-3} 
+ \vec{S}_{8r+1} \cdot \vec{S}_{8r+2} \right) \nonumber \\*
& & \qquad \qquad + J_2 \sum_r \vec{S}_r \cdot \vec{S}_{r+2}.
\label{eq:h57}
\end{eqnarray}
The legs are not equivalent. PBC ladders have inversion centers at every other 
site (3, 7, 11, \ldots) of the odd-numbered leg, and the unpaired spins at 
large $J_1$ are at these sites. Second order perturbation theory returns a F 
effective exchange $J_{\mathrm{eff}} \sim -J_2^2/J_1$ between the unpaired 
spins that are adjacent to the same end of a $J_1$ rung. The ladder has a F 
phase at large but finite $J_1$, albeit with small $J_{\mathrm{eff}}$. In 5/5 
ladders with $J_{\mathrm{eff}} > 0$, the unpaired spins are next to the 
opposite ends of a $J_1$ rung. 

\begin{figure}
\includegraphics[width=0.9\columnwidth]{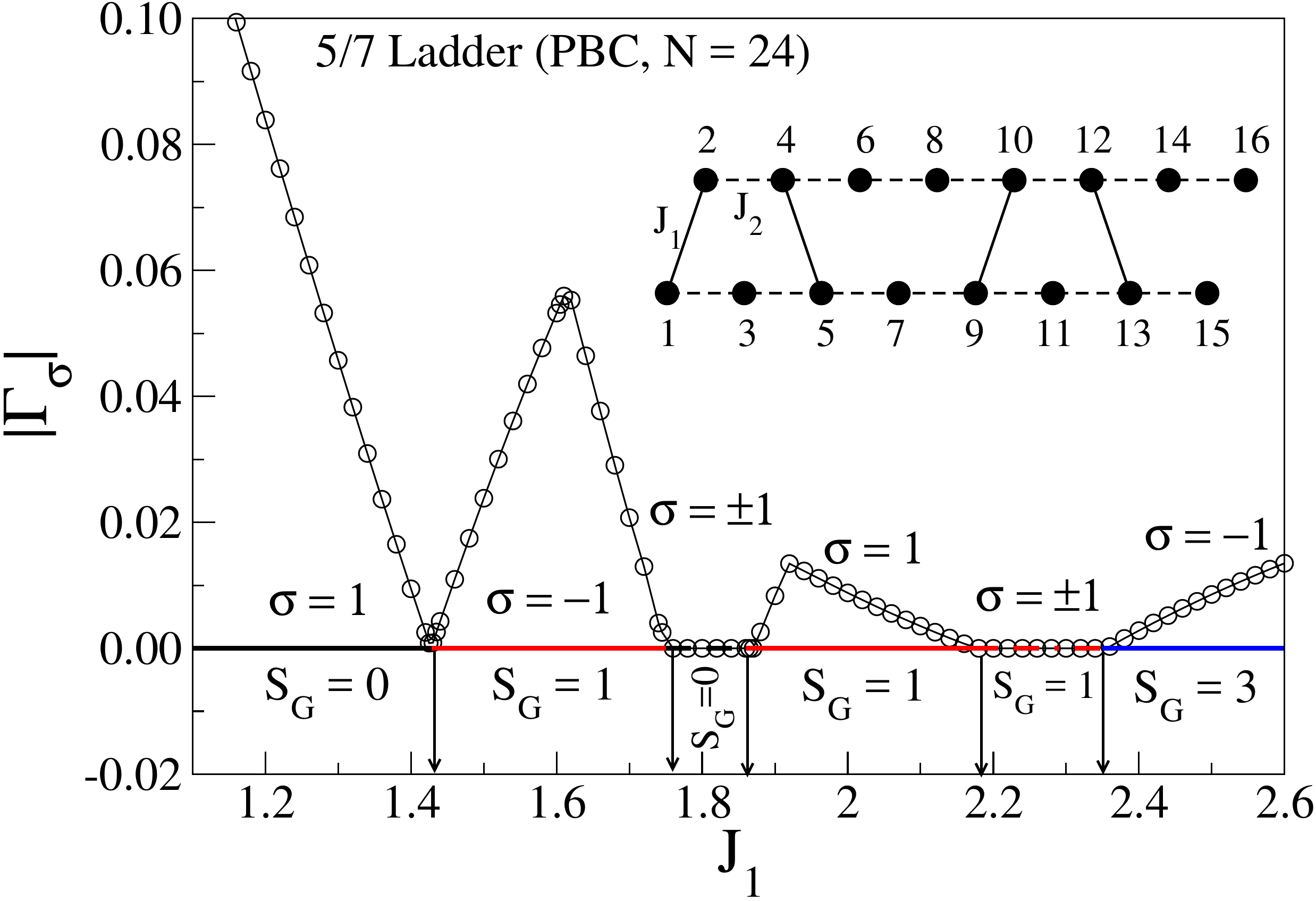}
\caption{\label{fig8}Excitation $\Gamma_{\sigma}$ in Eq.~\ref{eq:gma-sig} of a 
5/7 ladder with PBC and 24 spins; inset: two unit cells. The quantum phases 
have spin $S_G$ and inversion symmetry $\sigma$.}
\end{figure}

ED results for the PBC ladder of $N = 24$ spins are shown in Table~\ref{tab:57corl} 
and Fig.~\ref{fig8}. The sites are second, fourth and sixth neighbors in a leg. 
The $J_1 = 1$ spin correlations are close to the available thermodynamic results 
at $J_1 = 0$: $C_2 = 0.1820$ and $C_4 = 0.1040$. The GS is in the singlet sector 
at $J_1 = 1$ or 1.8 and in the $S_G = 3$ sector at $J_1 = 5$ or 40.
The signs of $C(3,11)$ and $C(3,15)$ are reversed at $J_1 = 1.8$. All the 
correlations approach $0.250$ at large $J_1$ as expected for an HAF with F 
exchange and spins at every fourth site.
\begin{table}
\caption{\label{tab:57corl}Spin correlation functions $C(j,k)$ at sites 
3, 7, 11, 15 of the 5/7 ladder with PBC and $N = 24$. Refer to Fig.~\ref{fig8}, 
inset, for site numbers.}
\begin{ruledtabular}
\begin{tabular}{ c  c  c  c}
$J_1$   &    $C(3, 7)$   &   $C(3, 11)$   &   $C(3, 15)$  \\  \hline
1	  & 0.1807   & 0.1161	 & 0.0974  \\
1.8	  & 0.1734   & -0.1102	 & -0.1747 \\
5	  & 0.2435   & 0.2372	 & 0.2389  \\
40	  & 0.2499   & 0.2498	 & 0.2498  \\ 
\end{tabular}
\end{ruledtabular}
\end{table}

As seen in Fig.~\ref{fig8}, the singlet GS is even under inversion ($\sigma = 1$) 
up to $J_1 = 1.43$, where it is degenerate with $\sigma = -1$, and it remains 
in the $\sigma = -1$ sector up to $J_1 = 1.76$. Then the singlet GS is doubly 
degenerate with $\sigma = \pm 1$ up to 1.87. The GS between 1.87 and 2.18 is a 
non-degenerate triplet in the $S_G = 1$, $\sigma = -1$ sector and a degenerate 
triplet from 2.18 to 2.35 with $\sigma = \pm 1$. The GS switches to $S_G = 3$ 
at $J_1 = 2.35$ at the onset of the F phase for three unit cells. The F phase 
of the $N = 16$ ladder and four unpaired spins is reached at the same $J_1$. 
DMRG results for longer OBC ladders of $N = 8n + 2 = 50$ or 100 spins confirm 
a F phase with $S_G = n$ for $J_1 > 2.35$ and $S_G/N = 0.125$ in the 
thermodynamic limit.

\begin{figure}
\includegraphics[width=0.9\columnwidth]{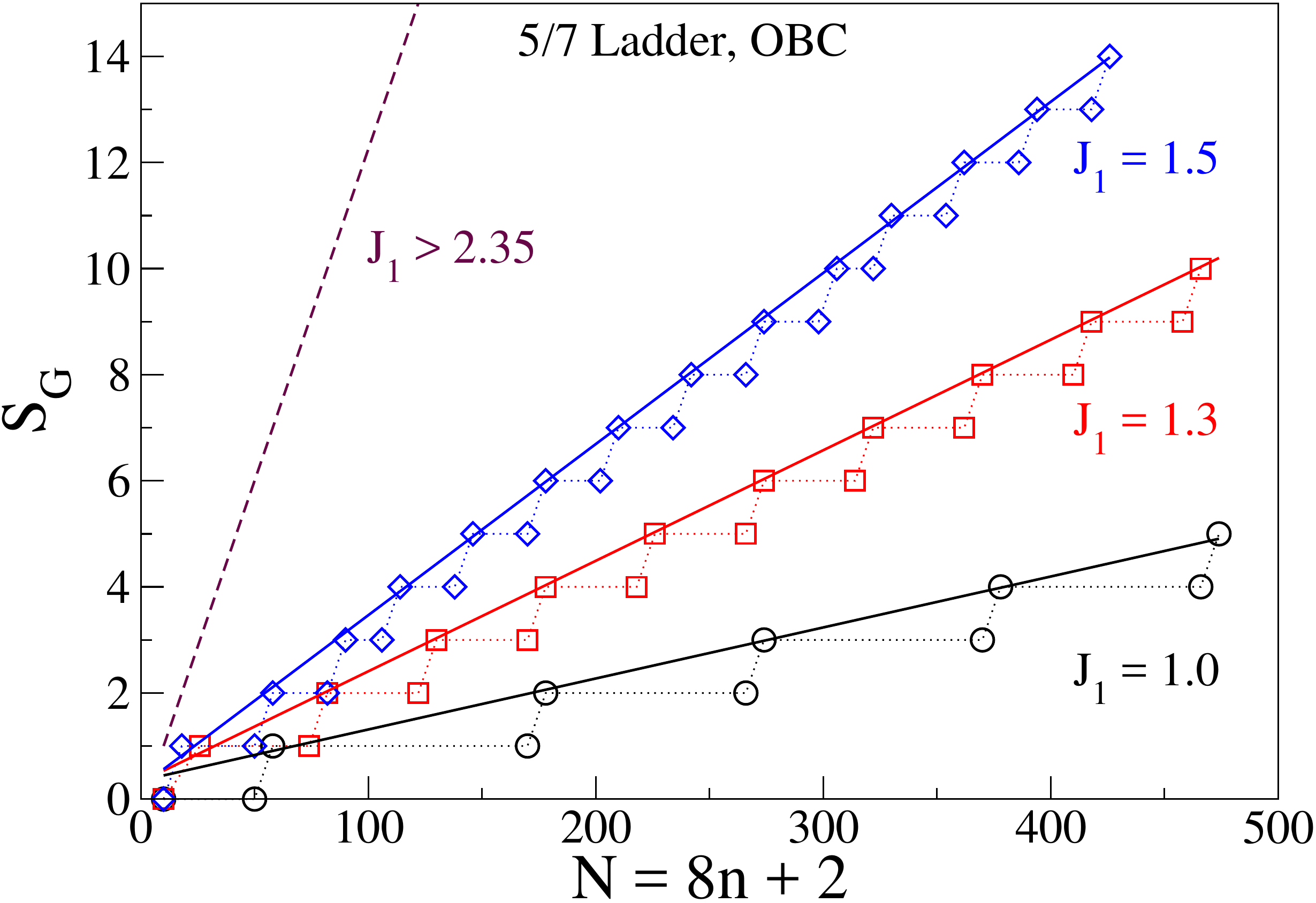}
\caption{\label{fig9}DMRG results for the size dependence of the ground-state spin 
$S_G(J_1,N)$ in 5/7 ladders with the indicated $J_1$. The dashed line is the F 
limit for $J_1 > 2.35$. The marked points correspond to system sizes at the edges 
of the plateaus are shown.}
\end{figure}

We return below to the multiple GS of small 5/7 ladders after reporting DMRG 
results for longer ladders. The F limit of one unpaired spin per ring is also 
reached at $J_1 = 2.35$ in long ladders with OBC as shown by the dashed line in 
Fig.~\ref{fig9}. The DMRG results for $S_G(J_1,N)$ vs. $N$ for smaller $J_1$ are 
reasonably linear in Fig.~\ref{fig9} and increase considerably more slowly. The 
location of $\Delta S_G(J_1,N) = 1$ steps is limited by the numerical accuracy 
of the total energy and minimally requires one unit cell, $\Delta N = 8$. 
Integer $S_G$ leads to constant plateaus over intervals in $N$, as shown best at 
$J_1 = 1$ where $S_G(1,N) = 5$ is reached around $N = 470$. There is roughly one 
unpaired spin per 12 rings (6 unit cells). ED for $N = 18$ or 26 indicates that 
the spins are delocalized at $J_1 = 1$ and cannot be rationalized by qualitative 
arguments. Aside from $S_G(1,N) = 1$ around $N = 54 \pm 4$, the $J_1 = 1$ plateaus 
in Fig.~\ref{fig9} have approximately equal widths implying $S_G \propto N$ that is consistent with a 
conjectured F phase in the thermodynamic limit. It is weak ferromagnetism at best, 
and longer ladders will be needed to verify the conjecture. The proportionality of 
$S_G(J_1,N)$ to system size is better realized at $J_1 = 1.3$ or $1.5$ in ladders 
with a F phase.

\begin{figure}
\includegraphics[width=0.9\columnwidth]{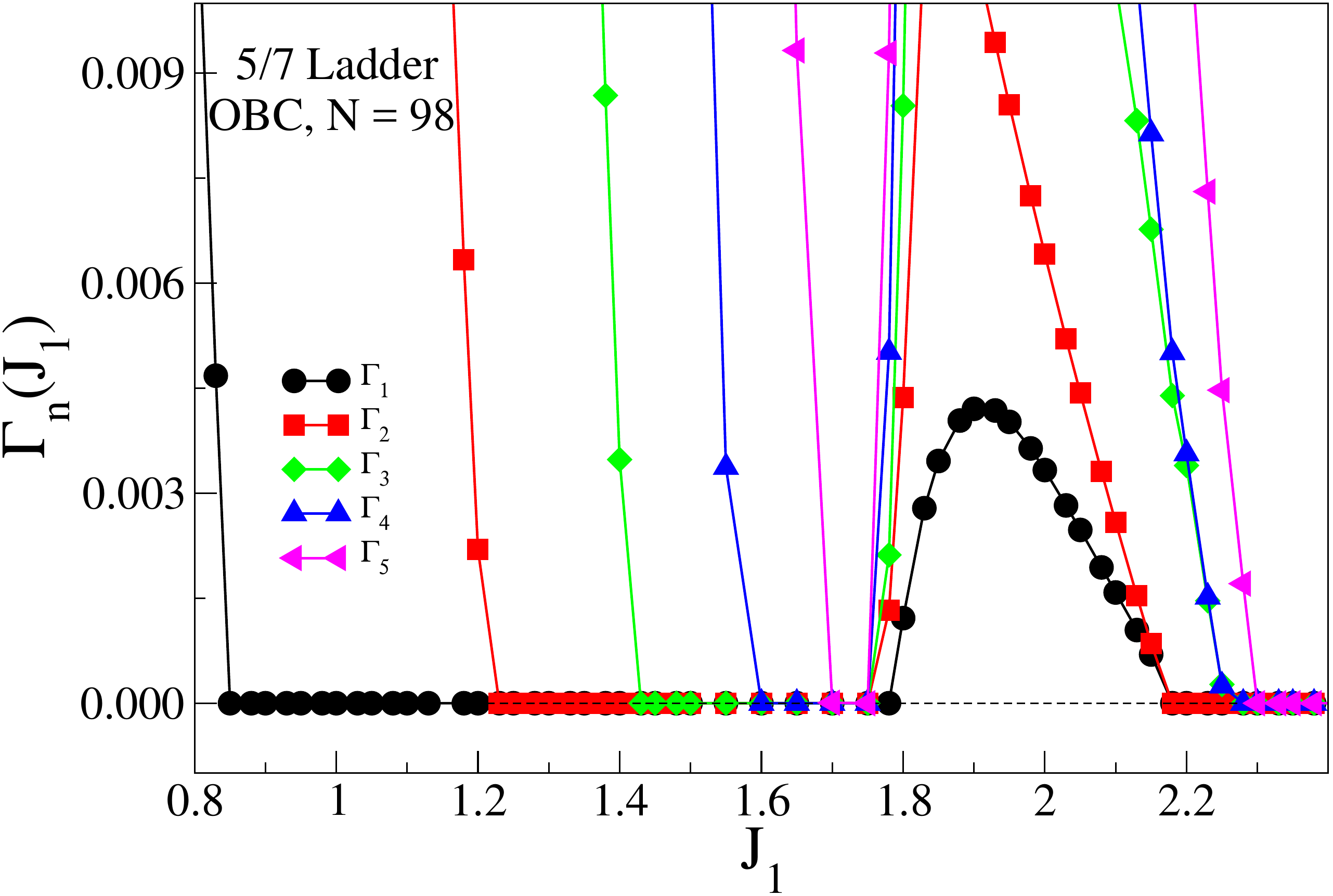}
\caption{\label{fig10}DMRG results for the GS spin $S_G(J_1,N)$ of a 5/7 
ladder of $N = 98$ spins and OBC as a function of $J_1$. The AF phase with $S_G = 0$ 
is regained between $J_1 = 1.75$ and 2.18.}
\end{figure}

We anticipated that $S_G(J_1,N)$ would increase faster with system size on 
increasing $J_1$ from 1.5 to 2.35, where the limit $S_G = n$ of one spin per ring 
is reached. That is not the case, however. The DMRG results in Fig.~\ref{fig10} show
that the GS of the OBC ladder of $N = 98$ spins (24 rings) has $S_G = 0$ up to
$J_1 = 0.85$, $S_G = 1$ from 0.85 to 1.23, $S_G = 2$ from 1.23 to 1.43,
$S_G = 3$ from 1.43 to 1.60, and $S_G = 4$ from 1.60 to 1.75. The GS is again a singlet, $S_G = 0$, in the range
$1.75 < J_1 < 2.18$. Almost exactly the same $S_G = 0$ range is obtained for an OBC 
ladder of 50 spins, while the $S_G = 0$ range for the PBC ladder of 24 spins in 
Fig.~\ref{fig8} is limited to $1.75 < J_1 < 1.87$, beyond which we have $S_G = 1$ 
up to $J_1 = 2.35$. The longer ladders do not have a triplet GS in this range. 
Increasing $J_1$ leads to a remarkable reentrant AF phase at intermediate 
$1.75 < J_1 < 2.18$. We do not understand how frustrated exchange and inequivalent 
legs in 5/7 ladders return a singlet GS at intermediate $J_1$.

The triplet GS in Fig.~\ref{fig8} is doubly degenerate ($\sigma = \pm 1$) in the 
$J_1$ interval 2.18 to 2.35. Finite 5/7 ladders with PBC and $N = 16$ or 24 spins have 
broken vector chiral symmetry in this range. The degeneracy is between $S^z = 1$, $\sigma = 1$ 
and $S^z = -1$, $\sigma = -1$. The spin current at sites $j,k$ is the matrix element 
$\pm \kappa (k,j)$
\begin{eqnarray}
\kappa_z (j, k) & = & \langle \Psi_{G}(-) | \vec{S}_j \times
\vec{S}_k | \Psi_{G}(+)\rangle \nonumber \\
& = & \frac{i}{2} \langle \Psi_{G}(-) | 
\left(S_j^{+} S_k^{-} - S_j^{-} S_k^{+} \right) | \Psi_{G}(+) \rangle,
\label{eq:spcur}
\end{eqnarray}
where $| \Psi_{G}(\pm) \rangle = | \mathrm{G}(S^z = \sigma = \pm 1) \rangle$.

The arrows in Fig.~\ref{fig1}(a) indicate the direction of spin currents. In 
finite regular chains, spin currents are absent for purely isotropic exchange without an  
applied magnetic field. Either anisotropic exchange~\cite{furukawa2012} or an applied 
field~\cite{hikihara2008} is typically required for broken vector chiral symmetry in these systems. 
However, we find that skewed ladders lead to nonzero spin currents even for isotropic exchange in the 
absence of applied magnetic field.   

\section{\label{sec6}Discussion and Summary}
We have obtained the quantum phases of frustrated 2-leg ladders with skewed rungs and found 
both F and AF phases at large $J_1/J_2$ when an unpaired spin is localized on 
every odd-membered ring. Perturbation calculations on odd ringed ladders give a simple pattern of 
effective exchanges $J_{\mathrm{eff}}$ between unpaired spins. Delocalization
within 3-membered rings leads to first-order corrections with $J_{\mathrm{eff}} \propto J_2$ 
between sites with finite spin density. The 3/4 ladder is F with $J_{\mathrm{eff}} < 0$ 
while the 3/5 ladder is AF with frustrated $j_1 < 0$ and $j_2 > 0$ in Eq.~\ref{eq:heff-35}.

The unpaired spin at large $J_1/J_2$ is localized at one 
site in 5 and 7-membered rings with $J_{\mathrm{eff}} \propto ± J_2^2 / J_1$. The 
unpaired spin in larger rings is delocalized over an odd number of sites, three 
sites for 9 or 11-membered rings. The unpaired spins of the 5/5 or 7/7 
are on opposite sides of a $J_1$ rung with $J_{\mathrm{eff}} > 0$. They are on the same 
side (in the same leg) in the 5/7 ladder with $J_{\mathrm{eff}} < 0$. The 5/7 and 
3/4/5/4 ladders have 8 spins per unit cell and are F and AF respectively at large $J_1$, while 
the 7/7 and 5/4 ladders with 5 spin per unit cell are AF and F. Unpaired spins on 
opposite sides of a $J_1$ rung have $J_{\mathrm{eff}} > 0$ but a second $J_1$ rung 
in a 4-membered ring changes the sign to $J_{\mathrm{eff}} < 0$. The same result 
holds for unpaired spins in the same leg: $J_{\mathrm{eff}} < 0$ when separated 
by a $J_1$ rung, $J_{\mathrm{eff}} > 0$ when separated by two rungs.

2-leg ladders with skewed rungs have inversion symmetry at some sites, rather 
than at all sites in the $J_1$-$J_2$ model, the 3/3 ladder. It is therefore not 
surprising to find BOW phases with $S_G = 0$ in the 5/5 or 3/5 ladder around 
$J_1 \sim  J_2$. The BOW amplitudes are more complicated than the dimer phase 
of the $J_1$-$J_2$ model because the unit cells contain several spins instead 
of just one. Longer 5/5 or 3/5 ladders with PBC than considered here may also 
support incommensurate phases.

At intermediate $J_1/J_2$, the 3/5 and 5/7 ladders have magnetic phases with $S_G > 0$ 
but considerably less than one unpaired spin per ring. The 3/5 ladder 
is AF for both small and large $J_1$ but is magnetic at intermediate $J_1$ from about 
2.3 to 6.9. The 5/7 ladder is F with an unpaired spin per ring for $J_1 > 2.3$,
weakly F for $J_1$ = 1 to 1.6, and quite remarkably AF between $J_1 \sim  1.75$ and 2.17. 
Both ladders have
inequivalent legs in addition to odd-membered rings. The evolution of $S_G$ with 
$J_1$ is not monotonic in either ladder. As for the 5/7 ladder with $J_1 = J_2$, 
the DMRG result in Fig.~\ref{fig9} is consistent with $S_G \propto N$ and 
suggests weak ferromagnetism that remains to be confirmed in considerably 
longer ladders.

A novel feature of 2-leg spin ladders with skewed legs is that both small and 
large $J_1$ correspond to extended systems, two HAFs when $J_1 = 0$ and a 
Heisenberg chain with $J_{\mathrm{eff}}$ of either sign for spins localized on 
odd-membered rings when $J_1 \gg J_2$. The conventional 2-leg ladder, $H_{4/4}$, 
has localized singlets at rungs when $J_1 \gg J_2$ and a continuous evolution 
with $J_1$ from gapless HAFs on legs to localized rungs. The evolution of 
$S_G(J_1)$ of ladders with skewed rungs is more complex and includes magnetic 
phases at intermediate $J_1$ before reaching an AF or F phase according to the 
pattern of $J_1$ rungs.
\begin{acknowledgments}
MK thanks DST for Ramanujan fellowship and computation facility provided under the
DST project SNB/MK/14-15/137. SR thanks DST for funding this work under various
programs.
\end{acknowledgments}

\bibliography{skewed_ref}
\end{document}